\documentclass[pre,twocolumn,showpacs]{revtex4}
\usepackage{graphicx}
\usepackage{amssymb,amsmath}
\usepackage{enumerate}

\newcommand{\be}{\begin{eqnarray}}
\newcommand{\ee}{\end{eqnarray}}
\newcommand{\D}{\mathrm{d}}

\begin{document}

\title{Singular features in noise-induced transport with dry friction}

\author{Adrian Baule$^1$\footnote{Present address: School of Mathematical Sciences, Queen Mary University of London, Mile End Road, London E1 4NS, UK} and Peter Sollich$^2$}

\affiliation{
$^1$Benjamin Levich Institute, The City College of New York, New York, NY 10031, USA\\
$^2$Department of Mathematics, King's College London, London WC2R 2LS, UK
}

\begin{abstract}

We present an exactly solvable nonlinear model for the directed motion of an object due to zero-mean fluctuations on a uniform featureless surface. Directed motion results from the effect of dry (Coulombic) friction coupled to asymmetric surface vibrations with Poissonian shot noise statistics. We find that the transport of the object exhibits striking non-monotonic and singular features: transport actually improves for increasing dry friction up to a critical dry friction strength $\Delta^*$ and undergoes a transition to a unidirectional mode of motion at $\Delta^*$. This transition is indicated by a cusp singularity in the mean velocity of the object. Moreover, the stationary velocity distribution also contains singular features, such as a discontinuity and a delta peak at zero velocity. Our results highlight that dissipation can in fact enhance transport, which might be exploited in artificial small scale systems.

\end{abstract}


\maketitle

\section{Introduction}

Many transport processes in physics and biology rely on the generation of directed motion in the absence of an externally imposed force gradient. In recent years much research effort has been devoted to understanding the physical principles underlying this kind of fluctuation induced motion, and to the question of how one can implement these principles in artificial systems on micro- and nanoscales \cite{Juelicher97,Astumian02,Reimann02,Haenggi09}. Unidirectional motion results quite generically from the combination of (i) \textit{non-equilibrium energy} input, e.g., due to chemical or mechanical fluctuations, and (ii) a spatial or dynamical \textit{asymmetry}. In the paradigmatic example of a Brownian ratchet \cite{Cordova92,Magnasco93}, thermal energy alone is not sufficient to generate a particle drift in a ratchet shaped potential due to the restriction imposed by the second law of thermodynamics, as was discussed already in the classical works by Smoluchowski, Feynman, and Huxley (see, e.g., \cite{Reimann02} and references therein). A non-zero drift results only when detailed balance is broken, which can be induced, e.g., by periodic switching between low and high temperature states, such that the Brownian particles can diffuse more easily over the potential barriers causing a net motion in a direction prescribed by the asymmetry.

In this letter we consider a transport model in which directed motion of an object is generated by coupling noisy asymmetric vibrations, taken specifically as Poissonian shot noise (PSN), with dry (or Coulombic) friction. Both asymmetry and non-equilibrium energy input are in this case due to the vibrations, but by itself this would not be sufficient to generate directed motion. In addition, the moving object has to exhibit a {\em nonlinear resistance} to the imposed force, leading to a hysteresis in the contact line that rectifies the asymmetric vibrations. This nonlinear response arises due to the dry friction between the object and the surface.

Our model thus combines two rather generic ingredients: on the one hand, PSN represents stochastic impulses that occur with a certain frequency and can be considered as a generalization of the usual Gaussian noise, to which it converges in an appropriate limit \cite{Feynman}. Dry friction, on the other hand, is ubiquitous in nature and plays a central role for diverse phenomena not only in physics and engineering, but also in biology and geology \cite{Persson,Urbakh04}. In our model, the interplay of friction and noise leads not only to a non-zero mean velocity of the object in the absence of a mean force input, but also to non-monotonic and singular features in the transport properties. By varying the dry friction strength three distinct modes of motion of the object can be induced: diffusive, directed and unidirectional. Strikingly, in the directed motion regime, the object's mean velocity increases when the dry friction coefficient {\em increases}, indicating that dissipation can in fact enhance transport. This highly counterintuitive effect might be exploited, e.g., in biological soft matter systems such as tethered vesicles or colloidal beads moving on lipid bilayers \cite{Yoshina03,Yoshina06,Wang09}.

\section{A model for directed motion due to dry friction and shot noise}

The velocity $v(t)$ of the solid object of mass $m$ is described in the reference frame of the surface by the Langevin equation
\be
\label{model}
m\,\dot{v}(t)=-\gamma v-\sigma(v)\Delta m+\xi(t),
\ee
where $\xi(t)$ denotes the random force. For simplicity, the discussion is reduced to one dimension. The friction between the two solids is expressed phenomenologically by two friction terms: a dynamic friction linear in $v$ with strength $\gamma$, and a dry friction $-\sigma(v)\Delta m$ with strength $\Delta$. The sign function $\sigma(v)$, defined as $\sigma(v)=+1,0,-1$ for $v>0,=0,<0$, respectively, is singular at $v=0$ and models the non-analytic behaviour observed in the dynamics of a solid object due to dry friction. In particular, the dry friction force in Eq.~(\ref{model}) can lead to complicated nonlinear stick-slip dynamics, where the object alternates in an oscillatory or chaotic way between sticking and sliding states \cite{Persson,Urbakh04}. We will sometimes find it useful to regularize the sign function by replacing it with a scaled hyperbolic tangent, where in the limit
\be
\label{tanh}
\sigma(v)=\lim_{\epsilon\to 0}\tanh(v/\epsilon),
\ee 
the singularity is recovered. Note that the strength of the dry friction is assumed directly proportional to the mass of the object, in accordance with the classical laws of friction dating back to the work by da Vinci, Amontons and Coulomb \cite{Persson}.

Models in the form of Eq.~(\ref{model}) have been investigated recently both theoretically and experimentally for the cases when the vibrations $\xi(t)$ are specified as asymmetric oscillations \cite{Eglin06,Buguin06,Fleishman07}, or as Gaussian white noise with zero mean \cite{deGennes05,Hayakawa04,Baule10,Baule11,Touchette10,Menzel10,Goohpattader09,Goohpattader10,Mettu10}. In the first case, the dry friction effectively rectifies the asymmetric oscillatory vibrations, leading to a directed motion of the object. This rectification is due to the induced sticking of the object to the surface, so that a threshold force is needed to move the object. Directed motion results when the vibrations do not overcome the threshold symmetrically, even when the total force is zero on average. In the case of Gaussian noise, no directed motion results unless an additional bias, such as a constant force, acts on the object. As a generalization of Gaussian noise and in order to investigate the influence of random spatial asymmetry, we consider Eq.~(\ref{model}) with $\xi(t)$ taken as Poissonian shot noise (PSN).

PSN is a mechanical random force, which is usually represented by a sequence of delta shaped pulses with random amplitudes $A$ \cite{Feynman}
\be
\label{PSN}
\xi(t)=\sum_{k=1}^{n_t}A_k\delta(t-t_k).
\ee
The waiting time between successive pulses is assumed to be exponentially distributed with parameter $\lambda$, where $\lambda$ is the rate at which pulses arrive. Then $n_t$, the number of pulses in time $t$, follows a Poisson distribution
\begin{eqnarray}
P(n_t=n)=\frac{(\lambda t)^n}{n!} e^{-\lambda t},
\end{eqnarray}
with mean $\lambda t$. When a pulse occurs, its amplitude $A$ is sampled, independently for each pulse, from a distribution $\rho(A)$. The mean and the covariance of $\xi(t)$ are then given by
\be
\left<\xi(t)\right>&=&\lambda\left<A\right>_\rho\\
\left<\xi(t)\xi(t')\right>-\langle \xi(t)\rangle\langle\xi(t')\rangle&=&\lambda\left<A^2\right>_\rho \delta(t-t'),
\ee
where $\left<...\right>_\rho$ indicates an average over the amplitude distribution $\rho(A)$. In the following we assume that the noise does not exert a net force on the object, which requires a zero mean noise $\left<\xi(t)\right>=0$. This can be achieved, e.g., by choosing a amplitude distribution with a vanishing mean. For an arbitrary $\rho(A)$, the mean of the noise can be subtracted by hand, i.e., $\xi(t)\to\xi(t)-\langle\xi(t)\rangle$. This case will be considered in more detail below.

The PSN Eq.~(\ref{PSN}) converges to Gaussian white noise when $\lambda\to\infty$ and $\left<A\right>_\rho\to 0$ with $\lambda \left<A^2\right>_\rho={\rm const.}$ \cite{Feynman}. In this limit, the model Eq.~(\ref{model}) thus converges to Brownian motion with dry friction \cite{deGennes05}. In the following, we focus on the non-Gaussian parameter regime of PSN and consider the stationary properties of the velocity, which is sufficient for an understanding of the transport properties of the model. Even in the Gaussian case, the time-dependent statistics can only be determined approximately using formal analytical methods \cite{Baule10,Baule11,Touchette10,Menzel10}. In Ref.~\cite{Cebiroglu10} a similar dry friction model has been studied, where the noise is given by an asymmetric Markov process with zero mean that switches between two states. This type of noise also induces directed motion but no singular features are reported.

The velocity process Eq.~(\ref{model}) can be expressed more simply after division by $m$ as
\be
\label{model2}
\dot{v}=-F(v)+\xi(t),
\ee
where the frictional forces are captured in
\be
\label{friction}
F(v)\equiv \frac{1}{\tau}v+\sigma(v)\Delta.
\ee
The time scale $\tau=m/\gamma$ is the inertial relaxation time. The noise $\xi(t)$ now represents stochastic kicks with the dimension of acceleration, which can formally be taken into account by an appropriate rescaling of the noise strength $A_0\to A_0/m$. Taking the noise average in Eq.~(\ref{model2}) and considering the steady state where $\left<\dot{v}\right>=0$ immediately leads to an expression for the mean velocity of the object:
\be
\label{meanvel}
\left<v\right>&=&-\Delta\tau\left<\sigma(v)\right>\nonumber\\
&=&\Delta\tau\left(\int_{-\infty}^0 p(v)\D v-\int_0^\infty p(v)\D v\right).
\ee
Here, $p(v)$ denotes the stationary distribution of the velocity, which has to be determined from the Kolmogorov-Feller equation associated with Eq.~(\ref{model2}) as discussed below. Eq.~(\ref{meanvel}) predicts that a positive mean velocity is induced when the total probability of observing a negative velocity is larger than the total probability of observing a positive one, which is somewhat counterintuitive, but is here a consequence of the particular form of $p(v)$. Moreover, Eq.~(\ref{meanvel}) is valid for generic noise sources $\xi(t)$ with zero mean \footnote{In Ref. \cite{Buguin06} a relation analogous to Eq.~(\ref{meanvel}) has been derived for asymmetric oscillatory vibrations $\xi(t)$. However, this relation is only valid approximately for small $\Delta$.} and shows explicitly that a non-zero mean velocity of the object results from:
\begin{enumerate}[(a)]
\item{{\em Inertia} (non-zero $\tau$).}
\item{The influence of {\em dry friction} (non-zero $\Delta$).}
\item{An {\em asymmetric} stationary velocity distribution.}
\end{enumerate}
Clearly, since the friction is symmetric, the asymmetry in the stationary distribution can only be induced by an asymmetric noise $\xi(t)$. The asymmetric fluctuations then lead to directed motion of the object due to the presence of the nonlinear dry friction. For comparison, in an overdamped system, asymmetric PSN has also been shown to induce a macroscopic current in symmetric periodic potentials \cite{Luczka95}.

The velocity distribution $p(v,t)$ of the velocity process Eq.~(\ref{model2}) can be derived from the Kolmogorov-Feller equation associated with the PSN \cite{Feller,Haenggi78}
\be
\label{KF}
\frac{\partial}{\partial t}p(v,t)&=&\frac{\partial}{\partial v}F(v)p(v,t)\nonumber\\
&&{}+\lambda \left(\left<p(v-A,t)\right>_\rho -p(v,t)\right).
\ee
Under the conditions of stationarity and a zero probability current one obtains
\be
\label{nonlocal}
F(v)p(v)=\int_{-\infty}^\infty G(v-v')p(v')\D v',
\ee
where the Green's function $G(v)$ has the Fourier representation (indicated by a $\sim$)
\be
\tilde{G}(k)=-i\lambda\left(\frac{\tilde{\rho}(k)-1}{k}\right).
\ee
The integral representation Eq.~(\ref{nonlocal}) indicates that the PSN induces a {\em non-local diffusion} of the object.

\section{Exact solution for one-sided shots}

An analytical solution for the stationary distribution of Eq.~(\ref{KF}) can be found when the amplitude distribution $\rho(A)$ is given by an exponential distribution
\begin{eqnarray}
\label{rho}
\rho(A)=\frac{1}{A_0}e^{-A/A_0},
\end{eqnarray}
where all amplitudes $A$ are assumed to be positive, i.e., the PSN that we consider is \textit{one-sided}. In order to obtain a noise with zero mean, we take for $\xi(t)$
\begin{eqnarray}
\label{xi}
\xi(t)= \sum_{k=1}^{n_t}A_k\delta(t-t_k)-\lambda A_0,
\end{eqnarray}
so that $\left<\xi(t)\right>=0$ and the covariance is $2\lambda A_0^2\delta(t-t')$. The noise thus consists of a random white shot noise part and a deterministic part that is equivalent to a constant negative drift force on the particle. Since the shot noise acts only in a one-sided fashion here, the noise $\xi(t)$ is strongly asymmetric, even though its mean value is zero by construction.

In the case of general values $\lambda$ and $A_0$ the stationary velocity distribution becomes strongly asymmetric, which is due to a lower cut-off for the possible velocity values at
\be
\label{cutoffs}
v_-=\left\{\begin{array}{l} (\Delta-\Delta^*)\tau\qquad,\qquad 0<\Delta<\Delta^*\\ \\0\qquad\qquad\qquad,\qquad \Delta\ge\Delta^*,
\end{array}\right.
\ee
where the critical dry friction strength $\Delta^*$ is just the average of the PSN
\be
\Delta^*\equiv\lambda A_0.
\ee
The lower bound in Eq.~(\ref{cutoffs}) can be understood as follows. The stochastic kicks $z(t)$ appear with rate $\lambda$ and only positive amplitudes. Between the kicks, the object will relax deterministically according to
\be
\label{veldet}
\dot{v}=-\frac{1}{\tau}v-\sigma(v)\Delta-\Delta^*,
\ee
which represents the motion of an object under the influence of dry friction and a constant force $-\Delta^*$. This equation has two different fixed points depending on $\Delta$. For $\Delta<\Delta^*$, the fixed point is at $v^*=(\Delta-\Delta^*)\tau$, which follows from setting $\dot{v}=0$ in Eq.~(\ref{veldet}) and solving for $v$. However, for $\Delta>\Delta^*$ there is always a net force opposite to the direction of motion and the fixed point is at $v^*=0$. Physically, the object will not be able to move with a negative velocity in this regime because the dry friction dominates the constant force and the object will always get stuck over time when relaxing from a stochastic kick.
 
Both regimes result in a non-zero mean velocity as will be discussed in the following. To summarize at this point, there are three distinct modes of motion of the object:
 \begin{enumerate}
 \item{{\em Diffusive} motion ($\left<v\right>=0$) when $\Delta=0$.}
 \item{{\em Directed} motion ($\left<v\right>>0$) when $0<\Delta<\Delta^*$.}
 \item{{\em Unidirectional} motion ($\left<v\right>>0$ and $v(t)\ge0$) when $\Delta\ge \Delta^*$.}
 \end{enumerate}
Diffusive motion also occurs for non-zero $\Delta$ in the Gaussian limit of the PSN, which has been discussed in Refs.~\cite{Baule10,Baule11,Touchette10,Menzel10}. For $\lambda$ and $A_0$ held constant, so that $\Delta^*={\rm const.}$, the transitions between the three regimes, and hence from diffusive to unidirectional motion, take place upon increasing the friction coefficient $\Delta$. In other words, the transport on the surface in fact improves for {\em increasing} dry friction. In all three regimes of motion, individual velocity trajectories reveal transitions between sticking and sliding states for non-zero $\Delta$, which is a characteristic feature of motion subject to dry friction (figures not shown). In the unidirectional motion regime, Eq.~(\ref{meanvel}) for the mean velocity must break down, since it would predict a negative mean even though the object's instantaneous velocity is never negative. The reason for the breakdown is the appearance of a delta-peak in $p(v)$ at $v=0$.

To explore these effects quantitatively, we consider the Kolmogorov-Feller equation~(\ref{KF}), which under our assumption of one-sided exponential shot noise becomes
\be
\frac{\partial}{\partial t}p(v,t)&=&\frac{\partial}{\partial v}\left[F(v)+\Delta^*\right]p(v,t)\nonumber\\
&&-\Delta^*\frac{\partial}{\partial v}\frac{1}{1+A_0\partial/\partial v}p(v,t).
\ee
Interestingly, the stationary state of this equation can be found exactly for arbitrary $F(v)$ using an operator inversion, leading to $p(v)\propto h(v)$ with \cite{VanDenBroeck83}
\be
\label{p_dist}
h(v)=\frac{\lambda}{F(v)+\Delta^*}\exp\left\{-\frac{v}{A_0}+\int^v_0\frac{\lambda}{F(v')+\Delta^*}\D v'\right\}.
\ee
The stationary distribution can therefore be determined not only for the piecewise linear friction Eq.~(\ref{friction}), but also for the smooth representation Eq.~(\ref{tanh}) of the $\sigma(v)$ singularity. This allows for a validation of the singular effects of the dry friction by a limit procedure. For $F(v)$ given by Eq.~(\ref{friction}), the exact expression for the integral in Eq.~(\ref{p_dist}) can be determined separately for $v>0$ and $v<0$. The solution is then
\be
\label{dist_ex}
h(v)=\left\{\begin{array}{l}\frac{\lambda}{\Delta^*+\Delta}\left(1-\frac{v}{v_+}\right)^{\lambda\tau-1} e^{-\frac{v}{A_0}},\quad v>0\\ \\ \frac{\lambda}{\Delta^*-\Delta}\left(1-\frac{v}{v_-}\right)^{\lambda\tau-1} e^{-\frac{v}{A_0}},\, v_-<v<0,\end{array}\right.
\ee
where $v_-$ is given in Eq.~(\ref{cutoffs}) and $v_+=-(\Delta+\Delta^*)\tau$ (so that $v_+$ is the hypothetical fixed point of Eq.~(\ref{veldet}) that one would obtain by replacing $\sigma(v)$ by $1$). For $\Delta=0$, i.e., in regime 1, $p(v)$ is identical for $v>0$ and $v<0$ as expected, because there is then no singular effect from the dry friction.

\section{Singular features}

\begin{figure}
\begin{center}
\includegraphics[width=8cm]{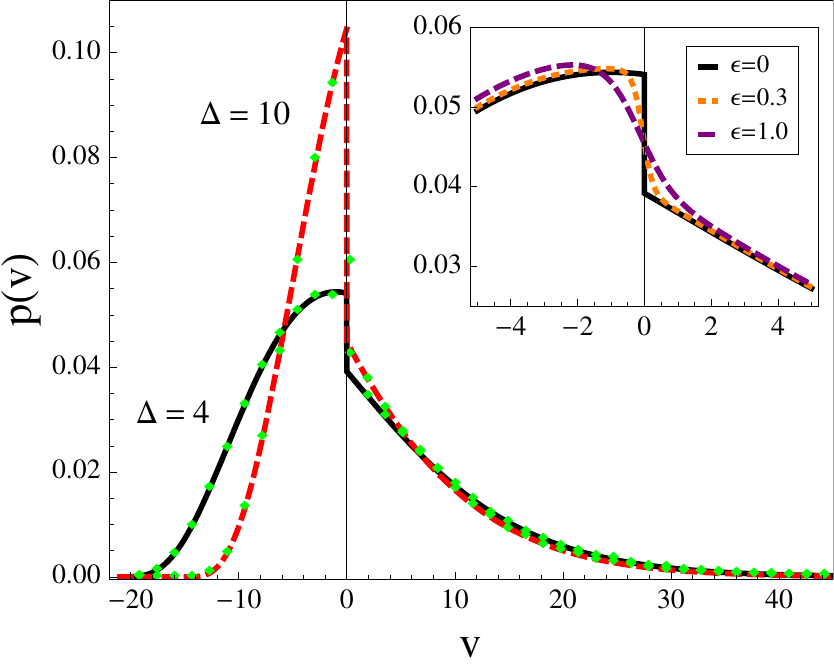}
\caption{\label{Fig_pstat}(Colors online) The normalized stationary velocity distribution Eq.~(\ref{dist_ex}) for two different values of $\Delta$ in the regime $0<\Delta<\Delta^*$. One notices the discontinuity in the distribution at $v=0$. The dots represent simulation results. Inset: the stationary velocity distribution using the $\tanh$ representation of the $\sigma(v)$ term, Eq.~(\ref{tanh}), and $\Delta=4$. Parameter values: $\lambda=5$, $A_0=5$, $\tau=1$.}
\end{center}
\end{figure}

Eq.~(\ref{dist_ex}) is plotted in Fig.~\ref{Fig_pstat} together with results from a direct simulation of the equation of motion (\ref{model}). The simulation uses a PSN increment method developed in \cite{Kim07} and the smooth $\tanh$ representation Eq.~(\ref{tanh}). For comparison, smooth versions of $p(v)$ are also shown in the inset. One clearly recognizes the trend towards a discontinuity at $v=0$ as $\epsilon\to 0$. The gap at $v=0$ satisfies
\be
\label{ratio}
\frac{p(0^-)}{p(0^+)}=\frac{\Delta^*+\Delta}{\Delta^*-\Delta},
\ee
and increases monotonically with $\Delta$ in the interval $0<\Delta<\Delta^*$. As $\Delta\to \Delta^*$ the gap becomes infinite while $v_-\to 0$, i.e., the support of $p(v)$ for $v<0$ vanishes. This indicates that the distribution exhibits a delta peak at $v=0$ for $\Delta= \Delta^*$. This peak in fact persists for $\Delta\ge \Delta^*$ as numerics show and can be obtained also from Eq.~(\ref{p_dist}), by considering a regularized version of $\sigma(v)$ as follows. For $\Delta>\Delta^*$ one can show that the lower cut-off $v_-$ on $v$ in such a regularized model is negative and of $O(\epsilon)$. The term $-v/A_0$ in the exponent in Eq.~(\ref{p_dist}) is negligible between this cut-off and $v=0$, and the total mass of $h(v)$ in this range becomes $1-\exp\{-\int_{v_-}^0\lambda/(F(v')+\Delta^*)\D v'\}$. The second term here vanishes because the integral is logarithmically divergent at the lower end. The distribution $p(v)$ thus contains finite probability mass within an $O(\epsilon)$ range of $v=0$, and this becomes a delta-peak for $\epsilon\to 0$.

The stationary solutions of diffusion processes are usually assumed to be continuous, which is required by the local nature of ordinary diffusion. In our case, PSN induces a non-local diffusion of the object, so that continuity of the solution is not required. For comparison, in the limit of white Gaussian noise $\xi(t)$ the Kolmogorov-Feller equation~(\ref{KF}) reduces to the standard Fokker-Planck equation and Eq.~(\ref{nonlocal}) would contain a term proportional to $p'(v)$ on the right-hand side. In this limit the stationary distribution can be found in a straightforward way \cite{Baule10}
\be
\label{pgauss}
p(v)\propto \exp\left\{-\frac{1}{D}\left(\frac{v^2}{2\tau} +|v|\Delta\right)\right\},
\ee
for a noise strength $D=\lambda A^2_0$. Eq.~(\ref{pgauss}) is continuous, but exhibits a cusp singularity at $v=0$. Due to the symmetry of $p(v)$ the mean velocity is zero in this case and the transport is purely diffusive.

\begin{figure}
\begin{center}
\includegraphics[width=8cm]{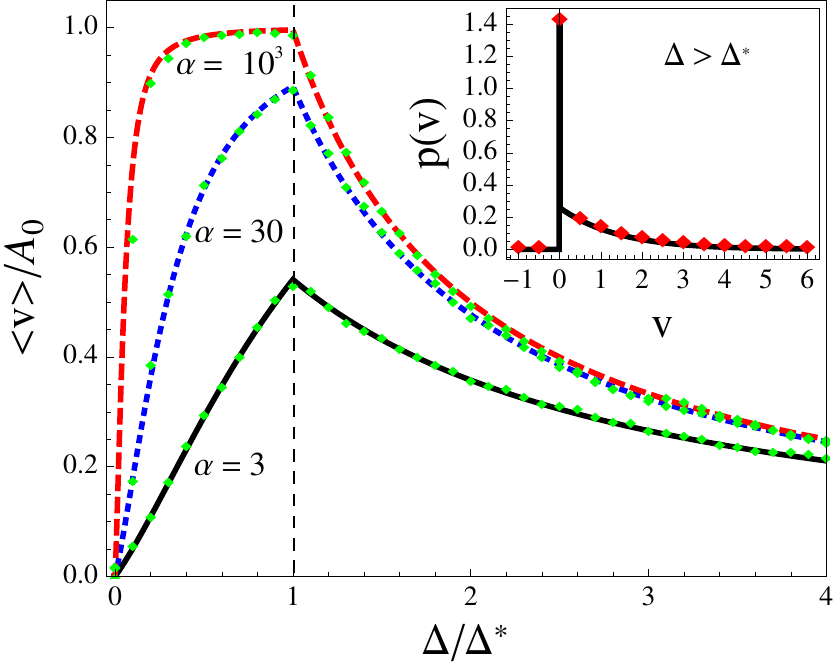}
\caption{\label{Fig_mvel1}(Colors online) Plot of the mean velocity $\left< v\right>$, given by Eqs.~(\ref{meanvel}) and (\ref{meanvel2}) for $0<\Delta<\Delta^*$ and $\Delta\ge\Delta^*$, respectively, as a function of $\Delta/\Delta^*$ for three different values of $\alpha=\lambda\tau$. At $\Delta=\Delta^*$ the crossover from the directed to the unidirectional regimes of motion leads to a cusp singularity. Inset: Plot of the stationary velocity distribution Eq.~(\ref{pstat_ge}) for $\Delta>\Delta^*$. The peak at $v=0$ stems from the delta-peak with amplitude $\Gamma_0$. The dots represent simulation results. Parameter values: $\Delta=1.5\Delta^*$, $\lambda=4$, $A_0=1$, $\tau=1$.}
\end{center}
\end{figure}

In order to understand quantitatively the gap in $p(v)$ as well as the appearance of the delta peak for $\Delta\ge \Delta^*$ we consider the integral equation~(\ref{nonlocal}). For the noise Eq.~(\ref{xi}) with exponentially distributed amplitudes, the Green's function is given by $G(v)=\Theta(v)\lambda e^{-v/A_0}$. Eq.~(\ref{nonlocal}) thus reads
\be
(F(v)+\Delta^*)p(v)=\lambda\int_{-\infty}^v e^{-(v-v')/A_0} p(v')\D v'.
\ee
Assuming first that $p(v)$ consists of two separate parts for $v> 0$ and $v<0$, but no delta-peak at $v=0$, the right-hand side is continuous and so the values of $p(v)$ either side of the gap at $v=0$ satisfy the relation
\be
\label{ratio2}
(\Delta+\Delta^*)p(0^+)=(\Delta^*-\Delta)p(0^-).
\ee
The ratio of the gap Eq.~(\ref{ratio}) follows immediately, if $0<\Delta<\Delta^*$. For $\Delta\ge\Delta^*$ the cut-off at $v^*_-=0$ enforces $p(0^-)=0$ so that Eq.~(\ref{ratio2}) would require $p(0^+)=0$. However, a zero value of $p(v)$ at $v=0^+$ is not consistent with the solution for $v>0$ in Eq.~(\ref{dist_ex}). The conclusion is that there has to be an additional contribution in order to satisfy Eq.~(\ref{ratio2}), which can only come from a delta peak in $p(v)$ at $v=0$. Therefore, we assume that, for $\Delta\ge\Delta^*$ (regime 3), the stationary distribution has the form
\be
\label{pstat_ge}
p(v)=\Gamma_0\delta(v)+\Theta(v)c_+h(v),
\ee
where $c_+$ is a normalization constant. Eq.~(\ref{ratio2}) is then replaced by
\be
(\Delta+\Delta^*)c_+h(0^+)=\lambda\Gamma_0,
\ee
so that together with the normalization condition $\int p(v)\D v=1$, the two unknowns $\Gamma_0$ and $c_+$ can be uniquely determined. This yields $c_+=\Gamma_0$ and $\Gamma_0=(1+\int_0^\infty h(v)\D v)^{-1}$. The distribution obtained in this way agrees well with simulation results (cf.\ inset of Fig.~\ref{Fig_mvel1}). The transition from the distribution Eq.~(\ref{dist_ex}) for $0<\Delta<\Delta^*$ to Eq.~(\ref{pstat_ge}) for $\Delta\ge\Delta^*$ occurs continuously, since the delta-peak amplitude $\Gamma_0$ is just the area of the $v<0$ part of $p(v)$ in Eq.~(\ref{dist_ex}) as $\Delta\to\Delta^*$. This follows from the limit: $\lim_{\Delta\to\Delta^*}\int_{v_-}^0 h(v)\D v= 1$.

In the case $\Delta\ge \Delta^*$, Eq.~(\ref{meanvel}) is not valid, but an analogous simple expression for the mean velocity can be derived using a limit procedure. Using the smooth $\tanh$ representation Eq.~(\ref{tanh}), the mean velocity of the process Eq.~(\ref{model2}) can be written in the form
\be
\left< v\right>&=&-\Delta \tau\int_{v_-}^\epsilon\tanh\left(\frac{v}{\epsilon}\right)h(v)\D v\nonumber\\
&&{}-\Delta \tau\int_{\epsilon}^\infty\tanh\left(\frac{v}{\epsilon}\right)h(v)\D v,
\ee
where $v_-$ is the lower cut-off as before. The second integral becomes $c_+\int_{0^+}^\infty h(v)\D v$ as $\epsilon\to 0^+$. For the first integral, one can show that
\be
\lim_{\epsilon\to 0^+}\int_{v_-}^\epsilon \tanh\left(\frac{v}{\epsilon}\right)h(v)\D v=-\Delta^*\Gamma_0,
\ee
when $\Delta\ge\Delta^*$. The analogue of Eq.~(\ref{meanvel}) is then
\be
\label{meanvel2}
\left< v\right>=\Delta^*\tau\Gamma_0-\Delta\tau c_+\int_{0^+}^\infty h(v)\D v.
\ee

In order to investigate the properties of the mean velocity we rewrite the expressions for $\left<v\right>$, Eqs.~(\ref{meanvel}) and (\ref{meanvel2}), in terms of the three parameters $A_0$, $\Delta/\Delta^*$, and $\alpha\equiv\lambda\tau$. This leads to the form $\left<v\right>=A_0 \,G(\frac{\Delta}{\Delta^*},\alpha)$, where $G$ has a different functional form for $0<\Delta<\Delta^*$ and $\Delta\ge\Delta^*$. The mean velocity rescaled by $A_0$ as a function of $\Delta/\Delta^*$ is plotted in Fig.~\ref{Fig_mvel1} for three different $\alpha$ values and shows a non-monotonic behavior. The crossover between the directed and unidirectional regimes of motion leads to a cusp singularity in $\left<v\right>$ at $\Delta=\Delta^*$. In the directed motion regime, $\left<v\right>$ increases with $\Delta$ up to a maximal value of $\left<v\right>=A_0$ as $\alpha\to\infty$, indicating that the transport generally improves on a rougher surface for the same noise characteristics (fixed $\lambda$ and $A_0$) in this regime. Optimal transport is achieved for $\Delta=\Delta^*$ and $\alpha\to\infty$. In other words, the object attains the maximal mean velocity when the mean of the noise balances the dry friction force (for $v>0$) and the object has no time to relax in between successive noise pulses. Eventually, in the unidirectional motion regime, $\left<v\right>$ decays to zero as $\Delta\to\infty$.

\section{Conclusion}

We have investigated an exactly solvable nonlinear model for the directed motion of an object due to dry friction and shot-noise with zero mean. The transport in this model behaves in a non-monotonic way as the dry friction strength is increased, exhibiting a transition to a unidirectional mode of motion at a critical dry friction strength $\Delta^*$. This transition is indicated by a cusp singularity in the mean velocity. The appearance of such a singularity at the transition between two qualitatively different modes of motion is reminiscent of a dynamical phase transition, which occurs here in a simple one-dimensional transport model.

A large variety of different distributions for the amplitudes of the shot noise can be incorporated in our model Eq.~(\ref{model}). In the presence of negative amplitudes, which can be realized, e.g., by choosing a shifted Gaussian or exponential amplitude distribution, the stationary distribution of the velocity could not be obtained in closed analytical form so far. However, some of the distinct singular features discussed here, namely the discontinuity and the delta peak in $p(v)$ at $v=0$ are a general consequence of the dry friction nonlinearity coupled to non-local shot noise. For this type of noise a path-integral solution similar to the one for Gaussian noise, derived in Refs.~\cite{Baule10,Baule11}, can also be developed. This will be discussed in a forthcoming article. 

The results presented could be tested experimentally in setups similar to the ones used by Chaudhury {\em et al} \cite{Goohpattader09,Goohpattader10,Mettu10}. Instead of Gaussian white noise vibrations one would have to induce vibrations with PSN statistics. By balancing the one-sided noise with a constant force, e.g., gravitation, a zero mean of the resulting force could be achieved. It would be interesting to observe in particular the sharp transition between the directed and unidirectional motion regimes. Moreover, the enhancement of transport for increased dissipation might be exploited in artificial or biological soft matter systems. Recent experimental results on diffusion statistics of tethered vesicles \cite{Yoshina03,Yoshina06} or colloidal beads moving on lipid bilayers \cite{Wang09} have been shown to agree with an effective dry friction model in the form of Eq.~(\ref{model}) \cite{Menzel10}. The PSN used in our model might be realized in these systems by an additional active chemical process that occurs on a separate time scale. To what extent the properties of our model persist in the presence of thermal noise remains to be investigated, but since thermal noise does not impose a force bias the qualitative features of the mean velocity should agree with our results.

\acknowledgments

AB gratefully acknowledges funding as a {\em Levich fellow}.

\end{document}